\RequirePackage{fix-cm}
\documentclass[twocolumn]{svjour3}          
\smartqed  
\usepackage{graphicx,xcolor}
\usepackage{amsmath,amssymb}

\begin{document}

\title{Avalanches in the athermal quasistatic limit of sheared amorphous solids: an atomistic perspective 
}


\author{C\'eline Ruscher        \and
       J\"org Rottler 
}


\institute{C\'eline Ruscher \at
             Institut Charles Sadron \\
		  23 rue du Loess
		  \\ F-67034 Strasbourg\\ 
		France \\
              \email{celine.ruscher@ics-cnrs.unistra.fr}           %
           \and
           J\"org Rottler \at
              Department of Physics and Astronomy and Quantum Matter Institute \\
              University of British Columbia \\
              Vancouver BC V6T 1Z1 \\
               Canada \\
              \email{jrottler@physics.ubc.ca}
}

\date{Received: date / Accepted: date}

\maketitle

\begin{abstract}
We study the statistical properties of the yielding transition in model amorphous solids in the limit of slow, athermal deformation. Plastic flow occurs via alternating phases of elastic loading punctuated by rapid dissipative events in the form of collective avalanches. We investigate their characterization through energy vs. stress drops and at multiple stages of deformation, thus revealing a change of spatial extent of the avalanches and degree of stress correlations as deformation progresses. We show that the statistics of stress and energy drops only become comparable for large events in the steady flow regime. Results for the critical exponents of the yielding transition are discussed in the context of prior studies of similar type, revealing the influence of model glass and preparation history.   

\end{abstract}

\section{Introduction}
\label{intro}

Granular materials, foams, metallic and colloidal glasses are yield stress materials. When subjected to deformation they initially respond elastically like a solid and flow when their intrinsic yield threshold is exceeded. In the athermal limit and at very slow driving, the flow shows sudden bursts of activity related to particle rearrangements, which materializes in the associated stress-strain curve as serrated, stick-slip type flow \cite{maloney2004,tanguy2006plastic,Bailey2007,Rodney_2011,dahmen2011simple,sun2010,antonaglia2014,denisov2016universality,OzawaPNAS}. \textcolor{black}{Microscopic instabilities that lead to sudden energy release also play an important role in determining the velocity dependence of the kinetic friction between solids \cite{mueser2002}. }
The distribution of the magnitude of the rapid stress releases, or avalanches of size $S$, exhibits scale free behaviour, $P(S) \sim S^{-\tau}$ characterized by an "avalanche exponent" $\tau$. In systems of linear dimension $L$, the distribution exhibits finite size effects and is truncated at $S_{c}\sim L^{d_f}$, where the exponent $d_f$ can be interpreted as a fractal dimension associated with the spatial extent of the dissipative events.   

Using large-scale atomistic simulations of amorphous packings in two and three dimensions, Salerno, Maloney and Robbins \cite{SalernoMaloneyRobbins2012,SalernoRobbins2013} performed a comprehensive characterization of the statistical properties of these avalanches. They considered both stress and potential energy release as indicators of the avalanche size, and computed the above mentioned exponents for different dynamical models that include overdamped and inertial dynamics in the steady state flow regime. Since their pioneering work, several new aspects have arisen that warrant revisiting the calculation of critical exponents of the yielding exponents through atomistic simulations. First, it was pointed out by Zhang et al. \cite{ZhangDamen2017} that stress and energy release might be only equivalent for large avalanches. Recently, Oyama et al. \cite{Oyama2020} suggested to fine-grain the distribution of avalanches into "precursors" and "mainshocks" that do not necessarily exhibit the same behavior. Lastly, Shang and Barrat \cite{ShangBarrat} pointed out that important information about the nature of the glassy state can be gained from avalanche statistics in the elastic regime, i.e. from the onset of deformation of a freshly quenched packing. This provides motivation to look beyond the steady state flow regime and to reexamine carefully the different definitions of avalanche size.  

In this contribution, we present result from atomistic simulations of simple shear of a 2D amorphous packing in the athermal quasistatic limit. We examine and compare in detail the distributions of stress and energy drops in three regimes of deformation, and find that their respective avalanche exponents only agree in steady state flow. Following ref.~\cite{Oyama2020}, we analyze separately precursors and mainshocks and confirm that the scale-free behavior is dominated by the precursor events. By studying the system size dependence of the stress fluctuations, we find that avalanches only become correlated in the post-yield flow regime and that these correlations influence the shape of $P(S)$.

\section{Simulations and methods}
\label{sec:1}
\subsection{Model}
We consider a 2D Lennard-Jones (LJ) glass-forming binary mixture where the $N_L$ large and $N_S$ small particles interact through the potential \cite{Lancon1988,FalkLanger1998,BarbotPatinet2018}
\begin{align}
U_{ab}(r) = \left \{
 \begin{array}{ll}
 \displaystyle
	4 \varepsilon_{ab} \left[ \left(\frac{\sigma_{ab}}{r} \right)^{12} - \left(\frac{\sigma_{ab}}{r} \right)^6 \right] + U_S,  \hspace{0.2cm} \forall \hspace{0.2cm} r \le r_{in} \\
	 \displaystyle \sum_{k=0}^4 C_k r^k \hspace{0.2cm} \forall \hspace{0.2cm},  r_{in} < r \le r_{cut} \\
 \displaystyle	0, \hspace{0.2cm} \forall \hspace{0.2cm} r > r_{cut}
	\end{array}
	\right .
\end{align}
where $\{a,b \}=\{L,S \}$ and $r$ is the distance between two particles.  The potential is shifted at the cutoff distance $r_{cut}=2.5 \sigma_{LS}$ and smoothed for $r_{in} < r \le r_{cut} $ where $r_{in}=2.0 \sigma_{LS}$ in order to ensure that $U_{ab}(r)$ is twice differentiable. 
The shift in energy $U_S$ and the coefficients $C_k$ are:
\begin{align}
U_S &= C_0 -  4 \varepsilon_{ab} \left[ \left(\frac{\sigma_{ab}}{r_{in}} \right)^{12} - \left(\frac{\sigma_{ab}}{r_{in}} \right)^6 \right] \\ \nonumber
C_0 &= -(r_{cut}-r_{in})[3C_1 + C_2(r_{cut}-r_{in})]/6 \\ \nonumber
C_1 &= 24 \varepsilon_{ab} \sigma^6_{ab} (r^6_{in} - 2\sigma^6)/r^{13}_{in} \\ \nonumber
C_2 &= -12 \varepsilon_{ab} \sigma^6_{ab} (7r^6_{in} - 26\sigma^6)/r^{14}_{in} \\ \nonumber	
C_3 &= -[3 C_1 + 4C_2(r_{cut}-r_{in}) ]/[3(r_{cut}-r_{in})^2]   \\ \nonumber	
C_4 &= [C_1 + C_2(r_{cut}-r_{in})]/[3(r_{cut}-r_{in})^3] 
\end{align}

The LJ parameters are $\sigma_{LL}=2 \sin(\pi/5)$, $\sigma_{LS}=1$, $\sigma_{SS}=2 \sin (\pi/10)$, $\varepsilon_{LL}=\varepsilon_{SS}=0.5$, $\varepsilon_{LS}=1$, and all masses are set to $m=1$. The ratio between large and small particles is chosen such as $N_L/ N_S = (1+ \sqrt{5})/4$, and we work at constant density $N/V=1.0206$. In what follows, the length, mass, energy and time units are expressed in term of $\sigma_{LS}$, $m$, $\varepsilon_{LS}$ and $\sigma_{LS}\sqrt{m/\varepsilon_{LS}}$ respectively. For this system, the glass transition temperature is $T_g=0.325 \varepsilon_{LS}/k_B$ where $k_B$ is Boltzmann's constant. To probe the system size dependence on the elasto-plastic observable, we consider linear system sizes ranging from $L=53$ to $L=400$.  
All systems are equilibrated at $T=0.65$ using Langevin thermostat with damping parameter $T_{damp}=1.0$ and then cooled down to $T=0$ at rate $dT/dt=2 \cdot 10^{-3}$.

\subsection{Deformation and detection of plastic events}

The systems are deformed following the athermal quasistatic shear (AQS) protocol. Simple shear is applied in the following way: an affine deformation is first performed by tilting the simulation box in the $x$-direction by an amount $\delta \gamma_{xy} L$, where $\delta \gamma_{xy}=10^{-5}$ is the strain increment, and then remapping the position of the particle inside the deformed box. In a second step, we allow the system to relax through an energy minimization using the conjugate gradient method. 

To detect plastic activity, we rely on an energy based criterion perfectly suitable for the AQS protocol \cite{LernerProcaccia2009,RuscherRottler2020}. Assuming harmonic approximation for the strain energy, we define an observable $\kappa$ that measures the difference between the energy associated to affine deformation $U_{\text{aff}}$ and the energy related to the inherent structure $U_{0}$ reached after relaxation:
\begin{align}
\kappa= \frac{U_{\text{aff}}-U_{0}}{N \delta \gamma^2}
\end{align}
If a plastic event takes place, the harmonic approximation does not hold anymore and $\kappa$ peaks.
In ref \cite{RuscherRottler2020}, we showed that in the elastic regime $\kappa \le  G_B/(2\rho)$ where $G_B \approx 26$ for our system is the Born shear modulus. Consequently, in what follow we assume that avalanches are associated with $\kappa \ge 30$.

Following \cite{Bailey2007}, we define the stress release during an avalanche as 
\begin{align}
   \Delta \sigma = \sigma_n - \sigma_{n+1} + G \delta \gamma
\end{align}
where $n$ is the number of AQS steps in chronological order and $G$ is the shear modulus calculated in the elastic regime.  Alternatively, we can also probe the energy drops due to plastic activity, which is given by
\begin{align}
\Delta E = E_n - E_{n+1} + V \sigma_n \delta \gamma.
\end{align}

In this work, we take the mechanistic point of view of experiments \cite{denisov2016universality} and consider stress drops as concrete indicators of particle rearrangements and therefore avalanches.

\subsection{Scaling relations}
In both transient and stationary regimes, stress drops compensate elastic loading in average leading to $\langle \Delta \sigma \rangle \sim  \langle  \Delta \gamma \rangle$ \cite{LinWyartPNAS}. Moreover, as the average distance to instability has finite size scaling, we define an exponent $\alpha$ that describes the system size dependence of the average stress drop,
\begin{align}
    \langle \Delta \sigma \rangle \sim L^{-\alpha} \hspace{0.2cm} \text{where} \hspace{0.2cm} \alpha>0
\end{align}

We define avalanches from the stress drops as $S=L^d \Delta \sigma$. As mentioned above, the avalanche distribution follows $P(S) \sim S^{-\tau} f(S/S_c)$ with $S_c \sim L^{d_f}$ and $f(S)$ a cutoff function.
This implies a scaling law linking these three exponents,
\begin{align}
    \alpha=d-d_f(2-\tau) \quad \text{for} \, \tau\ge 1.
    \label{alphatau}
    \end{align}

A further nontrivial exponent that enters the description of the yielding transition is the "marginality exponent" $\theta$, which describes the microscopic density of very unstable regions \cite{LinWyartEPL}. It is convenient to describe this density $P(x)$ in terms of the residual stress given by $x=\sigma_y-\sigma$, where $\sigma_y$ is a local yield stress of a region. In the extremal dynamics of the AQS protocol, one expects that the next failure event occurs at the weakest site $x_{min}$, and thus on average $\langle x_{min}\rangle\sim\langle \Delta \sigma \rangle$. The expectation value of the weakest site can be estimated using an argument from extreme value statistics,
\begin{align}
\int_0^{\langle x_{min}\rangle}P(x)dx\sim L^{-d}.
\end{align}
If one assumes that $P(x)\sim x^\theta$ for small $x$ then this implies $\alpha=d/(1+\theta)$ \cite{LinWyartEPL} and hence a second scaling law \cite{LinWyartPNAS},

\begin{align}
    \tau=2-\frac{\theta}{(\theta +1)}\frac{d}{d_f},
    \label{tauLW}
\end{align}
which provides the link between macroscopic stress drops and local weak zones in the system.

Both atomistic simulations \cite{RuscherRottler2020} and simulations of mesoscopic elastoplastic models \cite{Tyukodi2019,Ferrero2019}, however, have shown recently that $P(x)$ does not have this simple "pseudogap from" but becomes in fact analytic, i.e. a plateau develops in the limit $x\rightarrow 0$. Recently, Korchinski et al. \textcolor{black}{showed that the departure from the pseudogap regime originates from the far field effects associated with the mechanical noise created by spatially extended events \cite{Korchinski2021}. This deviation leads in the limit $x \rightarrow 0$ to a plateau whose entrance is directly related to the lower cutoff of the mechanical noise. This effect implies a modified scaling law for linking $\tau$ and $\theta$,
\begin{align}
    \tau=2-\frac{(d-d_f)\theta}{d_f}.
       \label{tauKRR}
\end{align}
This scaling law allows $\theta$ to be inferred from the avalanche size distributions.}

\section{Results}
\subsection{Stationary regime}

\begin{figure}
 \includegraphics[scale=0.5]{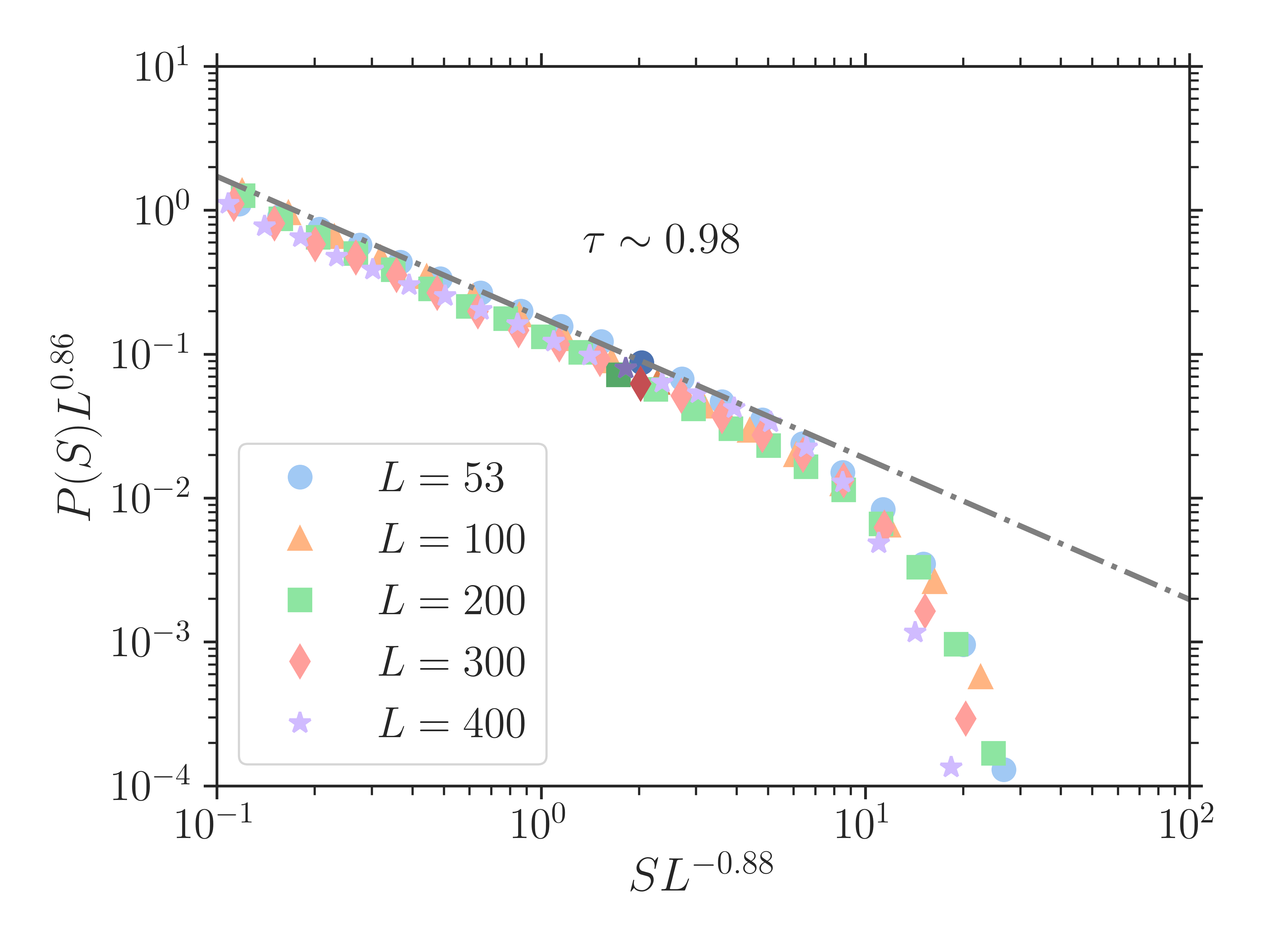}
  \includegraphics[scale=0.5]{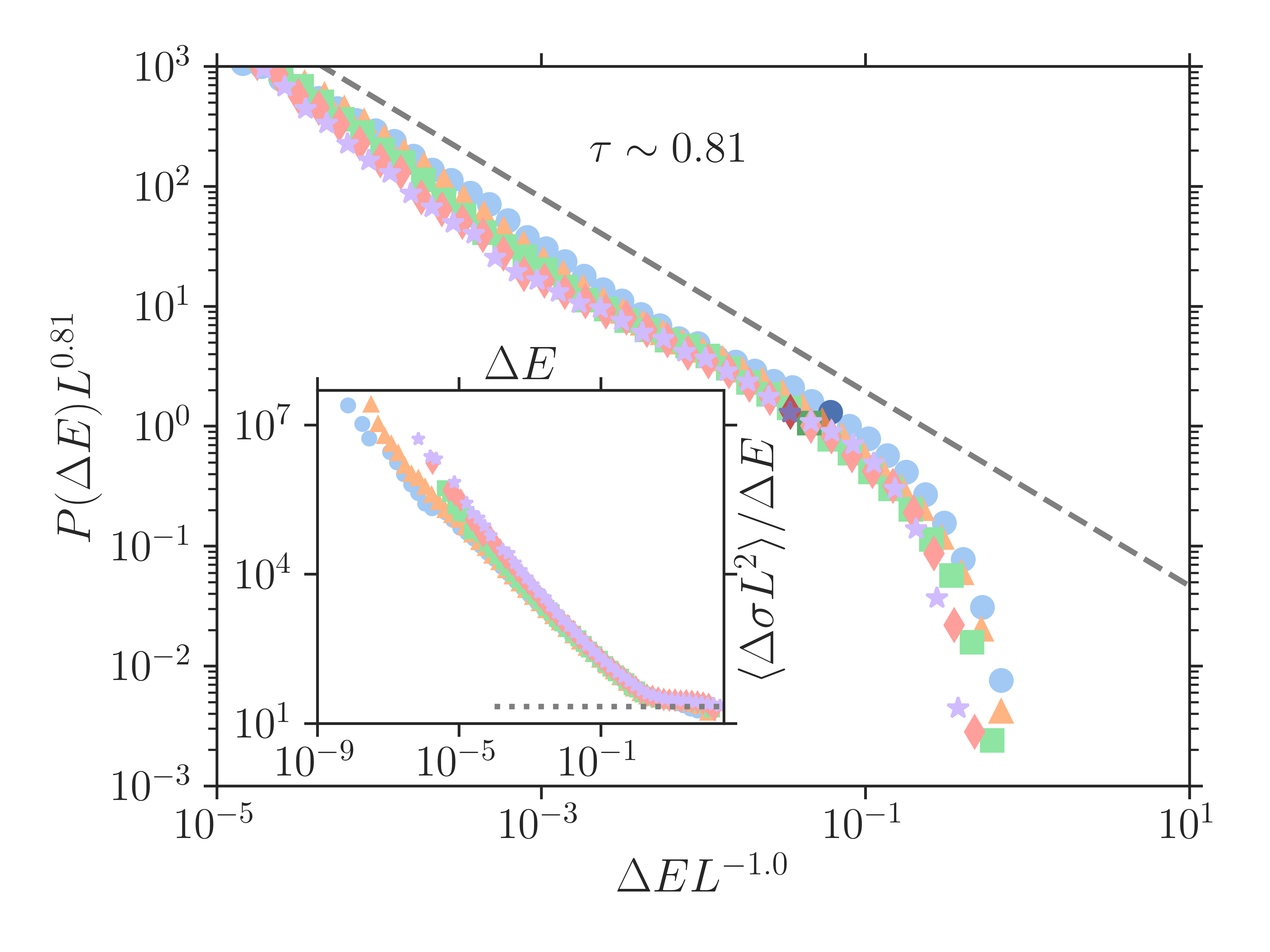} 
\caption{Top: Scaled distribution of avalanches $P(S)/P(S_c)$ vs $S/S_c$ with $S_c\sim L^{d_f}$computed from stress drops for the stationary regime. Bottom: Scaled distribution of energy drops $P(\Delta E)/P(\Delta E_c)$ vs. $\Delta E/\Delta E_c$ with $\Delta E_c\sim L^{d_f}$ in the stationary regime. The inset shows the ratio of window-averaged stress drops $\langle \Delta \sigma L^2 \rangle/\Delta E$ as a function of energy drops $\Delta E$. The dotted grey line indicates when stress and energy drops are proportional to each other. For both panels, the average values $\langle S \rangle$ and $\langle \Delta E \rangle$ are indicated with larger and darker symbols.}
\label{fig:1}       
\end{figure}

We compute in the stationary flow regime ($\gamma > 20 \%$) the avalanche distribution from stress and energy drops, $P(S)$ and $P(\Delta E)$, respectively.
In Figure \ref{fig:1}, we observe that these distributions exhibit the expected power law behaviour and a cut-off in the large avalanche limit.  \textcolor{black}{To extract the value of $\tau$ and $d_f$, we assume that the distributions can be described for each system size by a scaling function of the form $P(Y, L)=A S^{-\tau} \exp(-(Y/Y_c)^b)$ where $Y=S$ or $\Delta E$ \cite{Ferrero2019}.}
The avalanche exponents were computed as averages over the fit values from five different system sizes. We obtain \textcolor{black}{$\tau=0.98 \pm 0.03 $ for $P(S)$ but a smaller value $\tau=0.81 \pm 0.02$} for $P(\Delta E)$. The inset of Figure \ref{fig:1} (bottom) shows that, as already pointed out in ref.~\cite{ZhangDamen2017}, $S=\Delta \sigma L^2$ and $\Delta E$ are only proportional for large events. The difference in exponents can therefore originate from the nonlinear relation between $\Delta E$ and $S$.  

For both distributions, \textcolor{black}{we find the fractal dimension $d_f=0.88 \pm 0.10$ for $P(S)$ and $d_f=1.0 \pm 0.14$ for $P(\Delta E)$ from the finite size scaling of $S_c$ with $L$. While the values of $d_f$ are compatible with results reported in atomistic simulations, the avalanche exponents are lower than $\tau = 1.30$, $d_f= 0.90$  reported by Salerno and Robbins \cite{SalernoRobbins2013} and Liu et al. \cite{LiuBarrat2016} using atomistic simulations of two-dimensional amorphous solids with overdamped dynamics, as well as $\tau = d_f = 1.25$ reported by Zhang et al. with the AQS protocol \cite{ZhangDamen2017}}.  Interestingly, however, they agree much better with those obtained by Salerno and Robbins in what the authors call the "critically damped" regime ($\tau=1.0, d_f=0.8$) \cite{SalernoRobbins2013}.

\begin{figure}
 \includegraphics[scale=0.5]{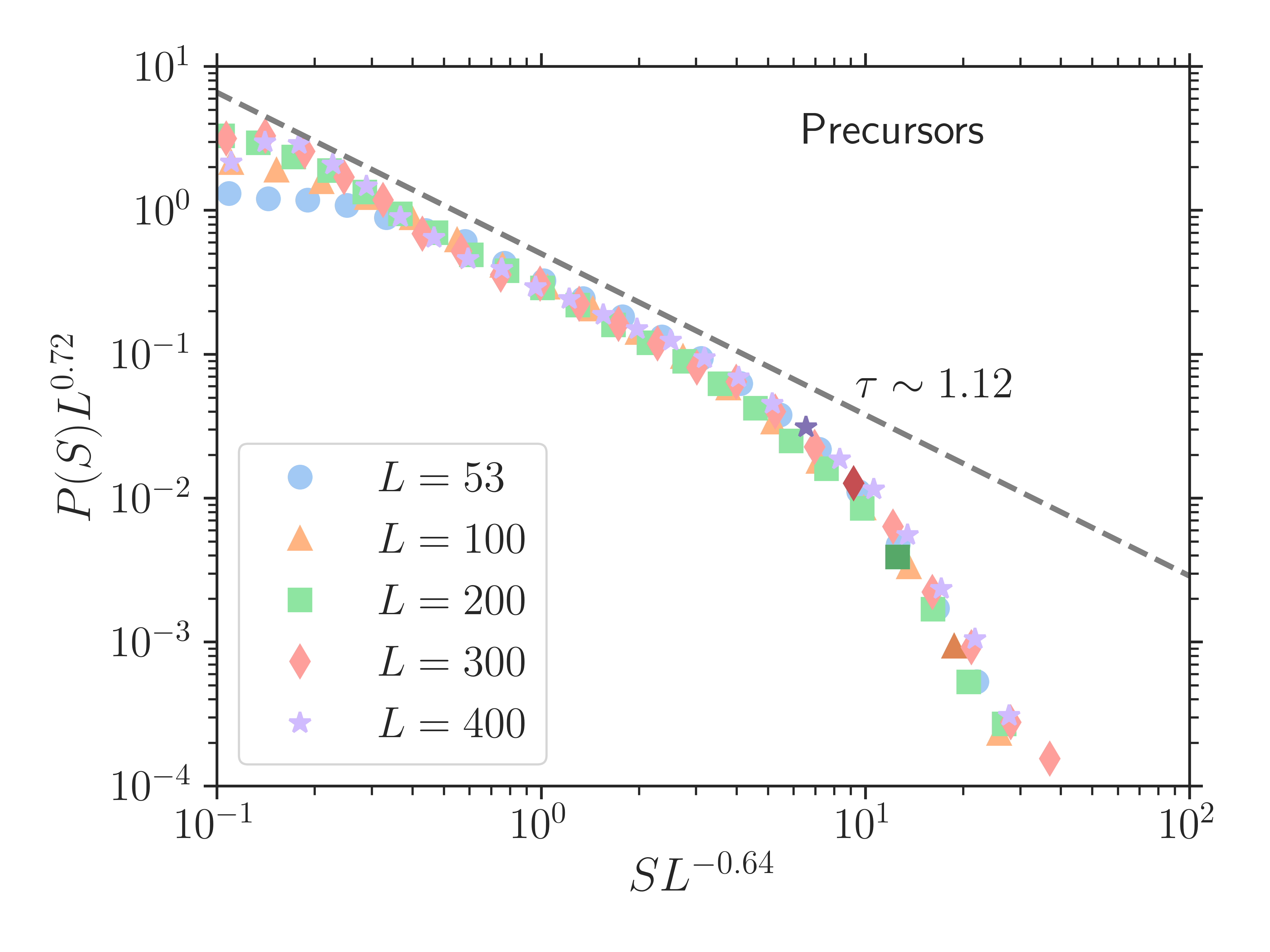}
  \includegraphics[scale=0.5]{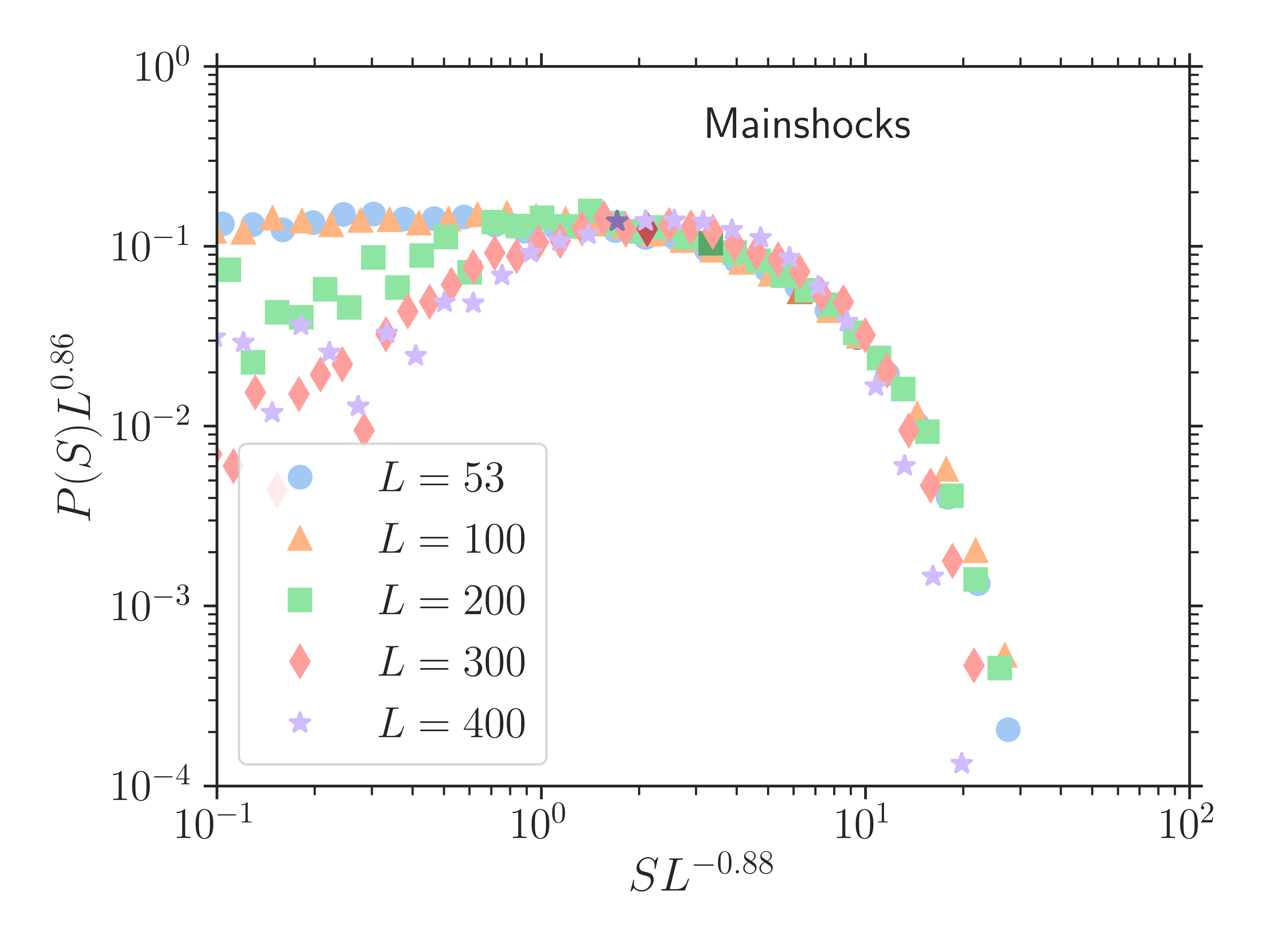} 
\caption{Rescaled distribution of avalanches in the stationary regime depending on whether they come from precursors (Top) or mainshocks (Bottom).  Average values $\langle S \rangle$ of the total ensemble of avalanches are indicated with larger and darker symbols.}
\label{fig:2}       
\end{figure}

In both distributions $P(S)$ and $P(\Delta E)$, we notice the presence of a bump in the large avalanche region. \textcolor{black}{More pronounced for $P(\Delta E)$, this bump has direct consequences of the value of $\tau$ as it is responsible for a flattening of the power law regime leading to two separate regions well visible for $P(\Delta E)$; in the small avalanches region $\tau$ is obviously larger than the value $0.81$. This comes from the fact that the fit is dominated by large avalanches.} Interestingly, the mean values of avalanche sizes $\langle S \rangle$ (and $\langle \Delta E \rangle$) are located in the bump region. A bump is not visible in the distributions obtained from overdamped dynamics \cite{SalernoRobbins2013,LiuBarrat2016} but becomes more prominent when the importance of inertia is increased \cite{SalernoRobbins2013}.

\begin{figure}
 \includegraphics[scale=0.5]{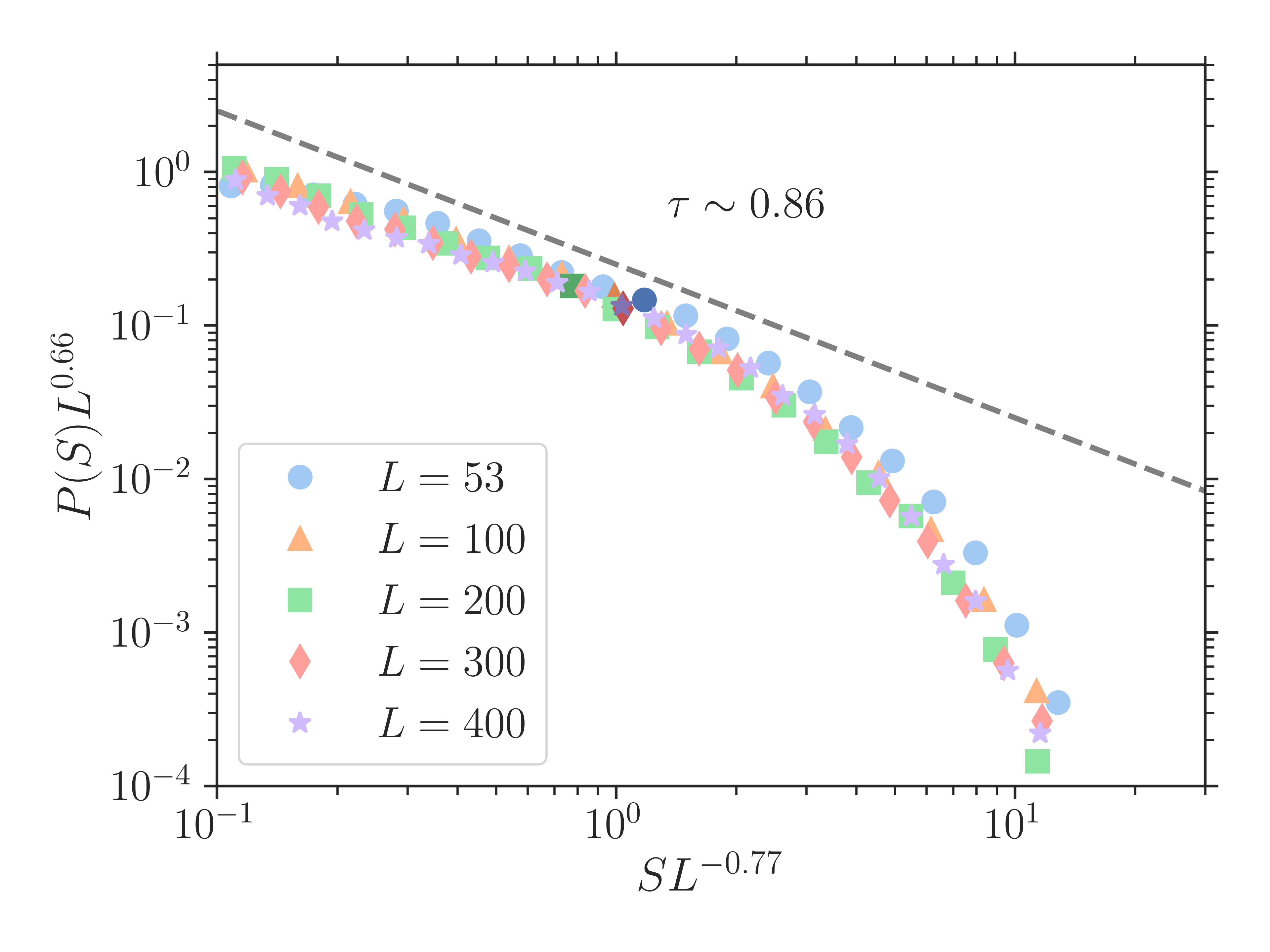}
  \includegraphics[scale=0.5]{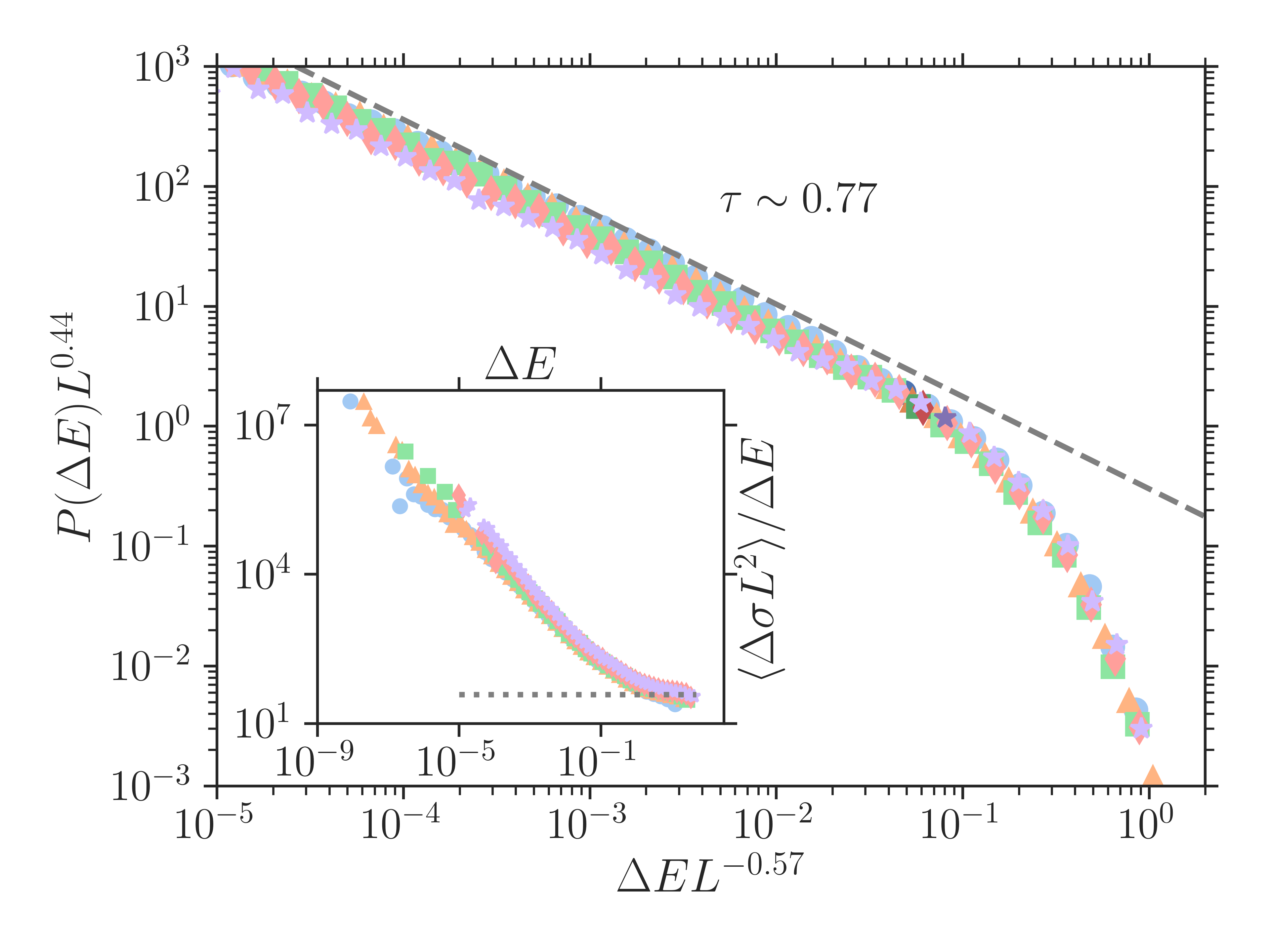}
 
\caption{Top: Scaled distribution of avalanches $P(S)$ computed from stress drops for the transient regime. Bottom: Distribution of energy drops $P(\Delta E)$ in the transient regime. The inset shows the ratio of window-averaged stress drops $\langle \Delta \sigma L^2\rangle/\Delta E$ as a function of energy drops $\Delta E$. The dotted grey line indicates when stress and energy drops are proportional to each other. For both panels, the average values $\langle S \rangle$ and $\langle \Delta E \rangle$ are indicated with larger and darker symbols.}
\label{fig:3}       
\end{figure}

\begin{figure}
 \includegraphics[scale=0.5]{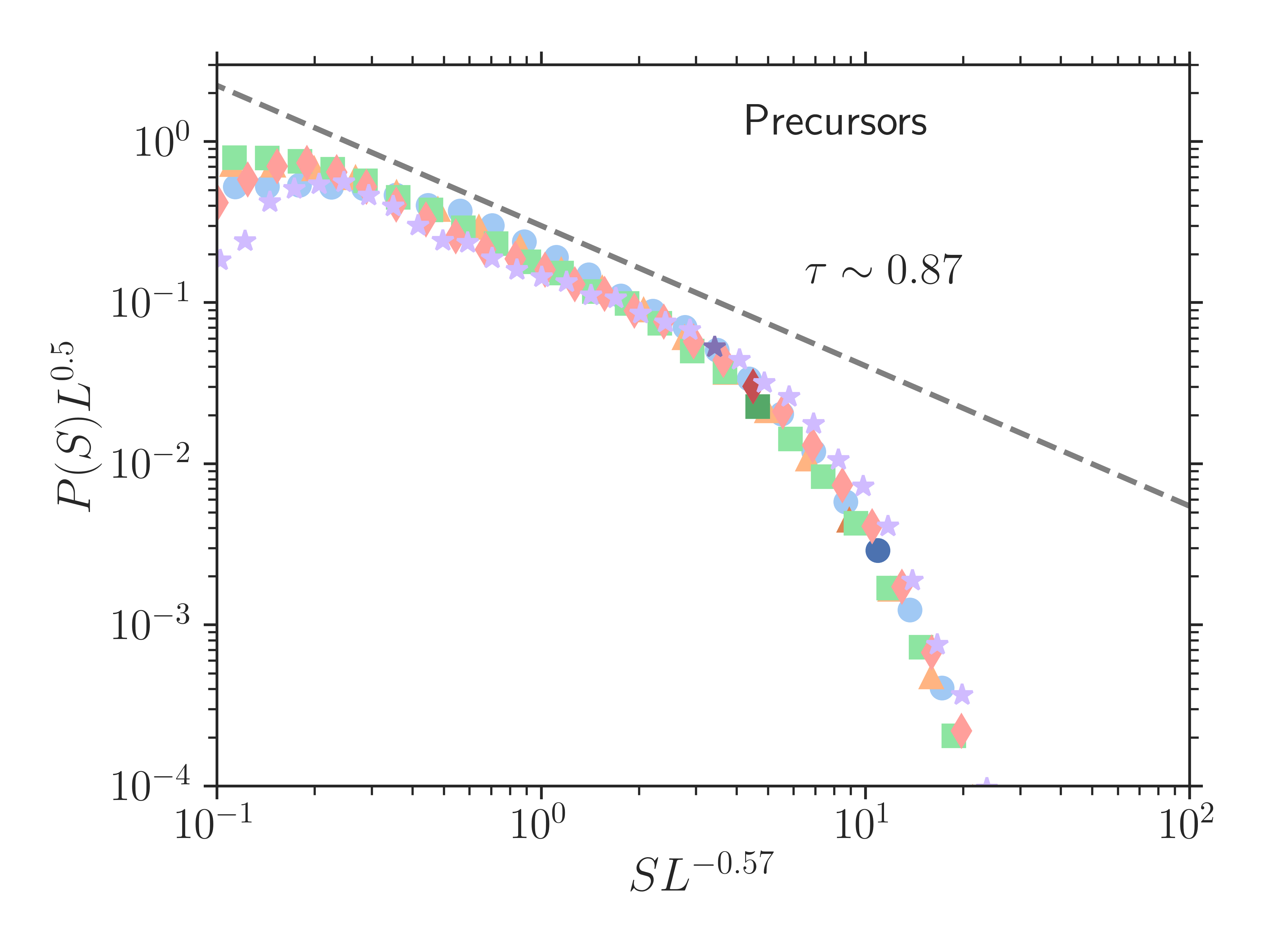}
  \includegraphics[scale=0.5]{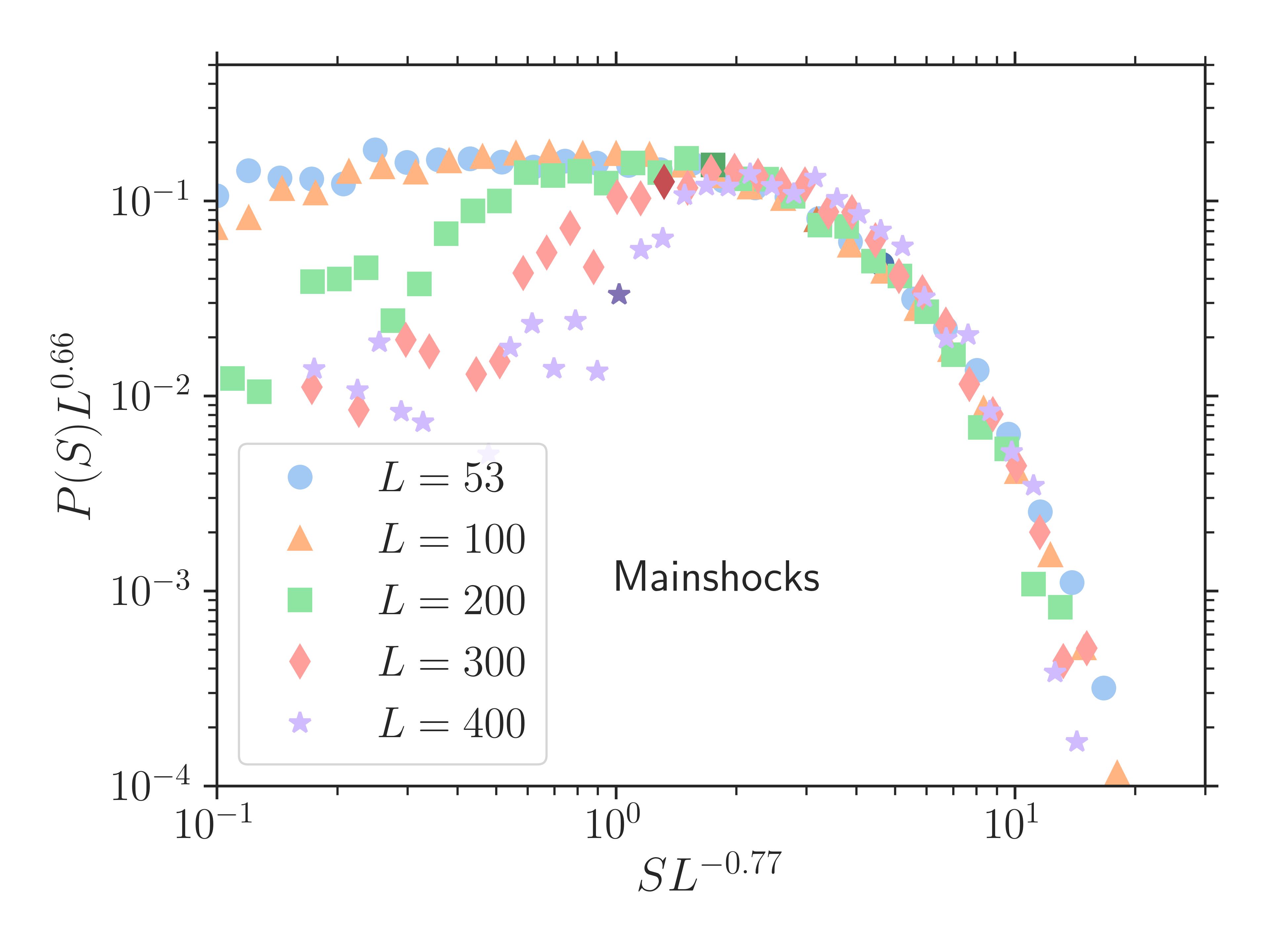} 
\caption{Rescaled distribution of avalanches in the transient regime depending on whether they come from precursors (Top) or mainshocks (Bottom).  Average values $\langle S \rangle$ of the total ensemble of avalanches are indicated with larger and darker symbols.}
\label{fig:4}       
\end{figure}

The precise mechanism responsible for the origin of the bump is still not well understood. In ref \cite{JaglaMF2015}, Jagla shows  for a mean-field model that the extremal dynamics \cite{Talamali2011} (uniform loading protocol) influences the shape of $P(S)$, leading to the appearance of a bump and a decrease in the value of $\tau$. However, if all values of $\tau$ in atomistic simulations and elasto-plastic models are significantly smaller than the mean-field value $\tau=1.5$, the appearance of a bump is not systematic, which questions the role of the protocol. More recently, Oyama and coworkers proposed to use a classification generally used for bulk metallic glasses to distinguish between two types of avalanches: precursors and mainshocks \cite{Oyama2020}.
While the displacement field of the mainshocks exhibits extended slip-band character, the precursors correspond to more localized events. They reported an avalanche exponent of $\tau=1.5$ for precursors while $\tau \approx 1.2$ when all avalanches were considered. They concluded that the decrease in $\tau$ is likely associated with mainshocks which are primarily responsible for the presence of a bump. The influence of spatial extent of avalanches on $\tau$ has also been reported in an elastoplastic model, where the mean field value of $\tau=1.5$ was found for point-like events while $\tau=1.37$ when all events where considered \cite{Korchinski2021}.

Following this idea of classifying avalanches, we compute the difference in the stress right before two consecutive avalanches $i$ and $i+1$, i. e. we look at the sign of $\sigma_n(i+1)-\sigma_n(i)$. Avalanche $i$ is defined as a precursor (resp. a mainshock) if the difference is positive (resp. negative). The rescaled avalanche distributions conditioned on precursors and mainshocks are shown in Figure \ref{fig:2}, where we observe the presence of a power law regime for precursors only. It is interesting to notice that the average value $\langle S \rangle$ is not located in the power-law region, meaning that only small plastic events are responsible for the scale free behaviour. As in ref.~\cite{Oyama2020}, the avalanche exponent \textcolor{black}{$\tau=1.12 \pm 0.07$ from precursors is larger than when all events are considered,} but smaller than the mean-field value $\tau=1.5$ reported by these authors \cite{Oyama2020}. 

\textcolor{black}{The fractal dimension decreases from  $d_f=0.88 \pm 0.10$ to $d_f=0.64 \pm 0.15$ when precursors are considered. This trend makes sense given that the precursors are events that are spatially less extended. For the mainshocks, the absence of a power law regime prevents us from directly fitting the scaling function  mentioned above. However, we find in Fig.~\ref{fig:2} that using $\tau$ and $d_f$ extracted from the whole distribution results in a good finite size scaling collapse of the mainshocks distribution.} Thus one can conclude that large plastic events modify the shape of the avalanche distribution by flattening it and consequently decreasing the apparent avalanche exponent. 

\begin{figure}
 \includegraphics[scale=0.5]{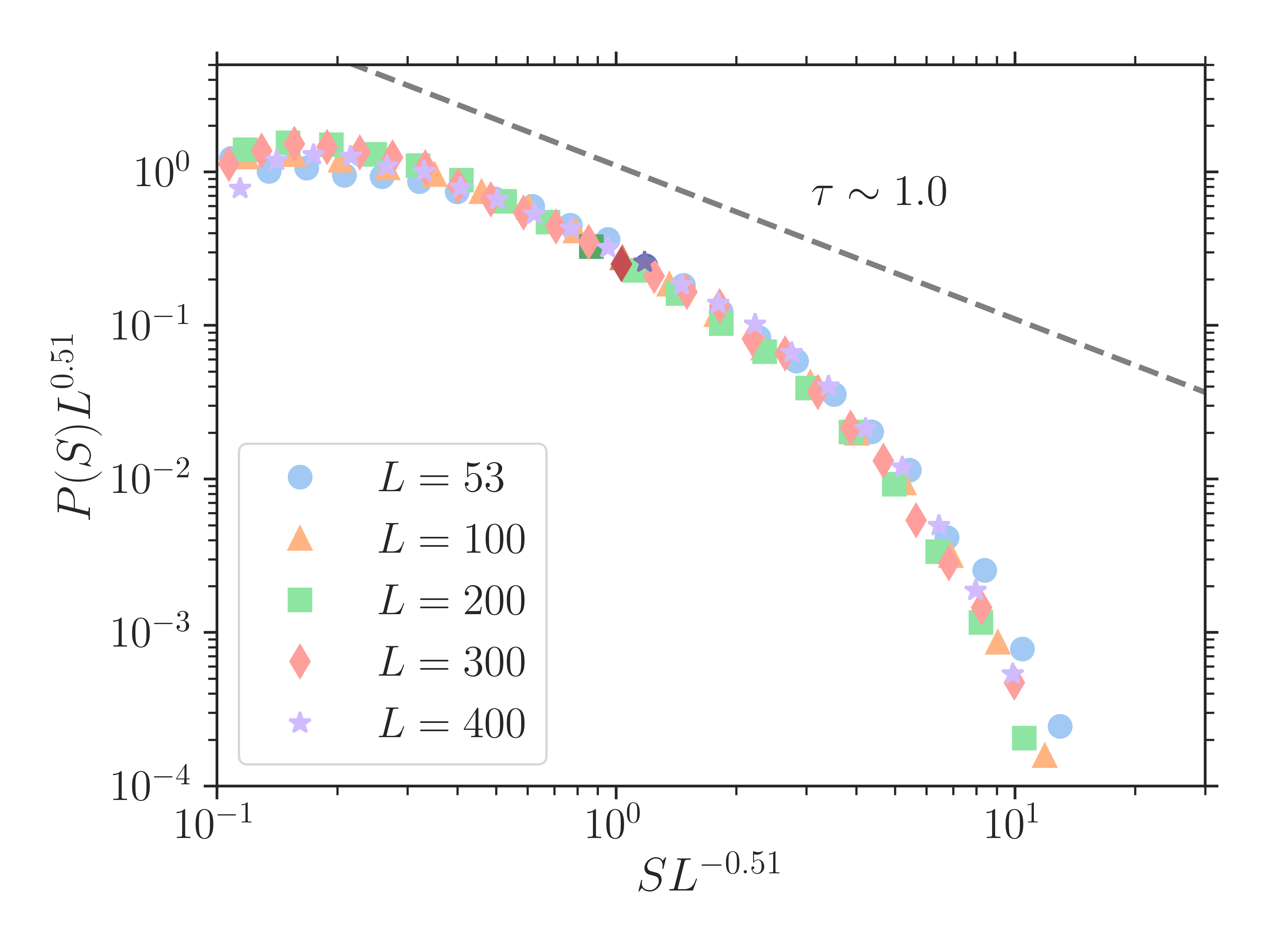}
  \includegraphics[scale=0.5]{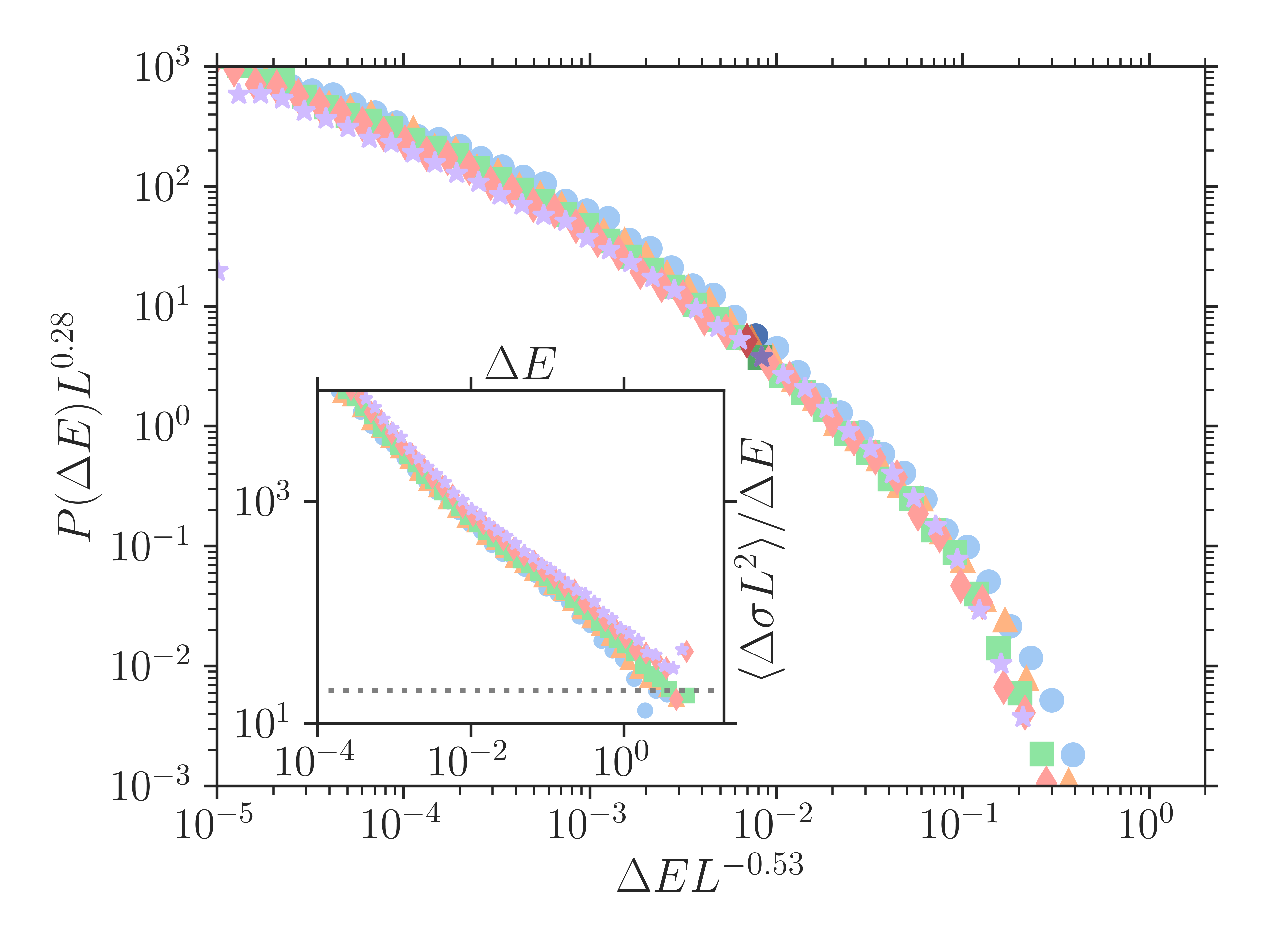}
 
\caption{Top: Scaled distribution of avalanches $P(S)$ computed from stress drops for the elastic regime. Bottom: Distribution of energy drops $P(\Delta E)$ in the elasic regime. The inset shows the ratio of window-averaged stress drops $\langle \Delta \sigma L^2\rangle/\Delta E$ as a function of energy drops $\Delta E$. The dotted grey line indicates when stress and energy drops are proportional to each other. For both panels, the average values $\langle S \rangle$ and $\langle \Delta E \rangle$ are indicated with larger and darker symbols.}
\label{fig:7}       
\end{figure}

\subsection{Transient regime}
Using the same methodology, we now investigate the transient regime (strain interval 2-4\%). The distributions of avalanches shown in Figure \ref{fig:3} computed from stress and energy drops reveal unique power-law regimes (with no bump) characterized by smaller values of the avalanche exponent \textcolor{black}{ and fractal dimensions, $\tau=0.89 \pm 0.04$, $d_f=0.77 \pm 0.15$ and $\tau=0.77 \pm 0.08$, $d_f=0.57 \pm 0.20$ for $S$ and $\Delta E$, respectively, than in the steady-state flow. As in steady state, however, the avalanche exponent given by the energy drops is smaller than that given by the stress drops due to the nonlinearities between $\Delta E$ and $S$ (see inset).}
It is interesting to note that for $P(\Delta E)$, the finite size scaling is better in the large $\Delta E$ region, confirming the important role of nonlinearities in the small avalanches region. 

To compare with the stationary regime, we again separate the avalanches into precursors and mainshocks. In the transient regime, the stress grows linearly with the strain, and therefore we expect more precursors than mainshocks on average. As as result, the distributions of precursors should be very similar to the total distributions. This is indeed what we observe in Figure \ref{fig:4} where we see that \textcolor{black}{the avalanche exponent does not vary as we find $\tau=0.87 \pm 0.13$. As in the stationary regime, the fractal dimension decreases, $d_f=0.57 \pm 0.16$.} In this case, the average value of avalanches $\langle S \rangle$ are located - with the exception of the smallest system - at the end of the power law regime of the precursors distributions or alternatively in the region where finite size scaling fails for the mainshocks distributions, suggesting that large avalanches are not playing a dominant role \textcolor{black}{regarding the avalanche exponent but influence the cutoff region in $P(S)$ as shown by the reduction in the fractal dimension.} 

\begin{figure}
 \includegraphics[scale=0.5]{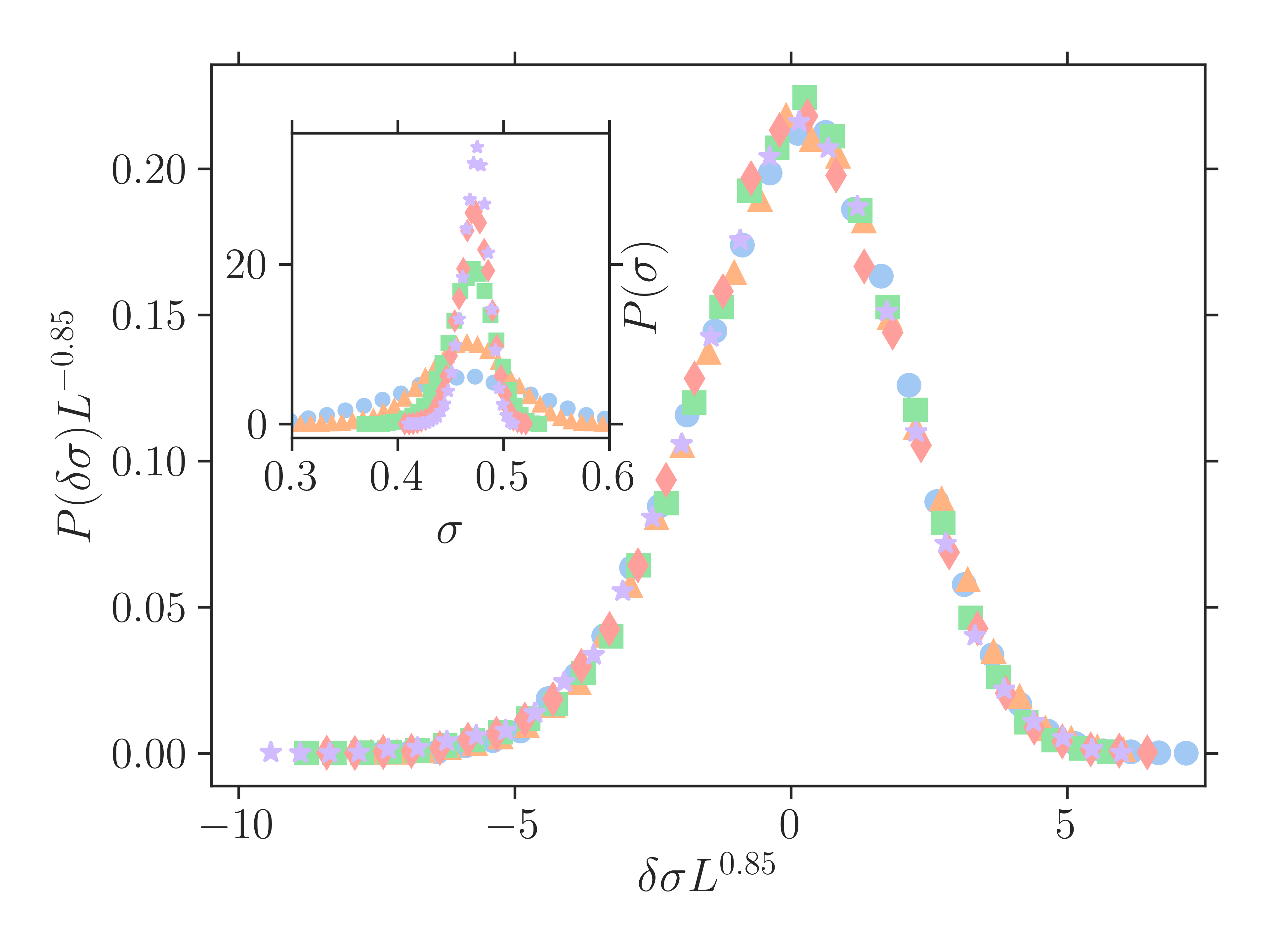}
  \includegraphics[scale=0.5]{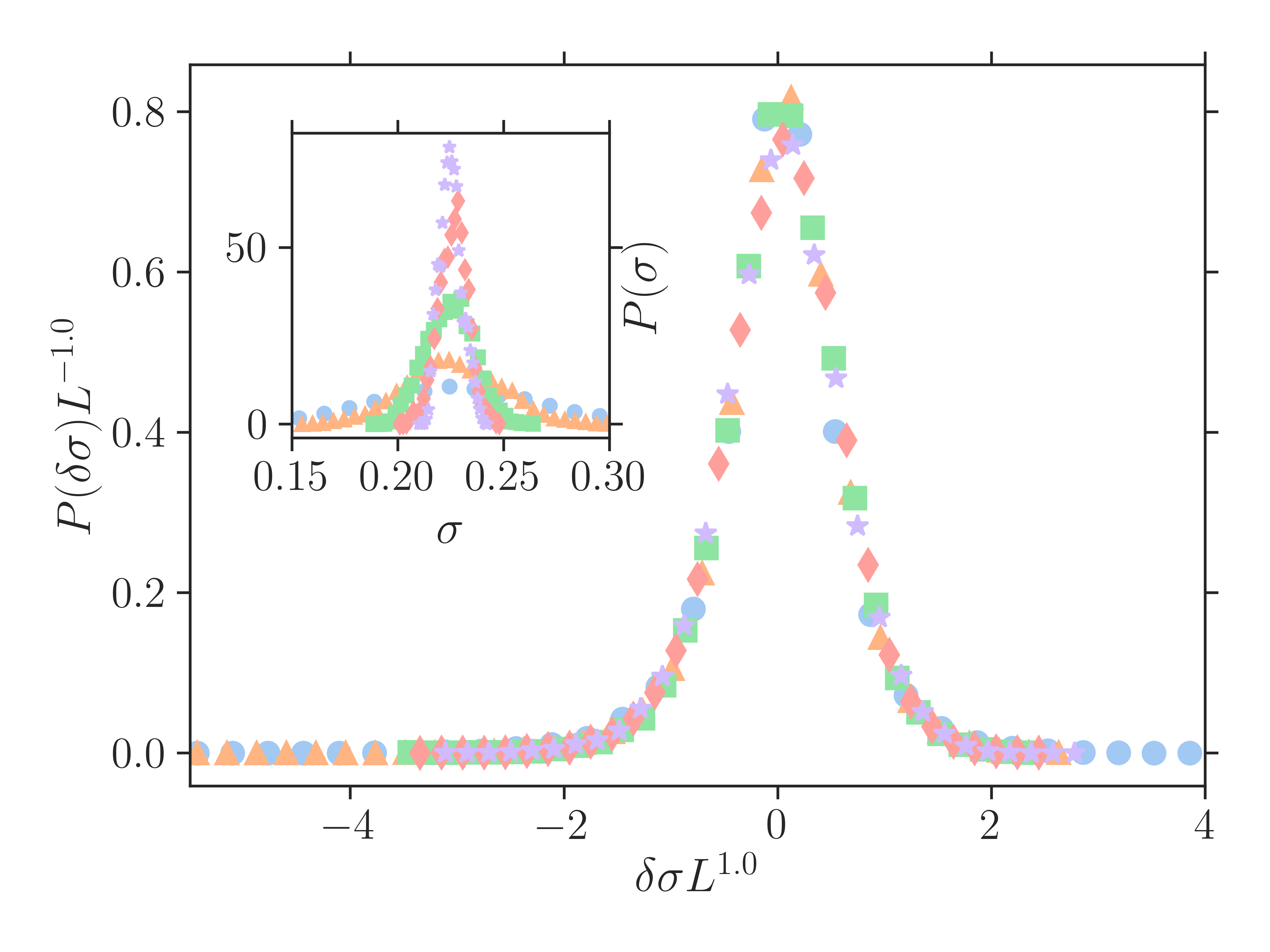} 
  \includegraphics[scale=0.5]{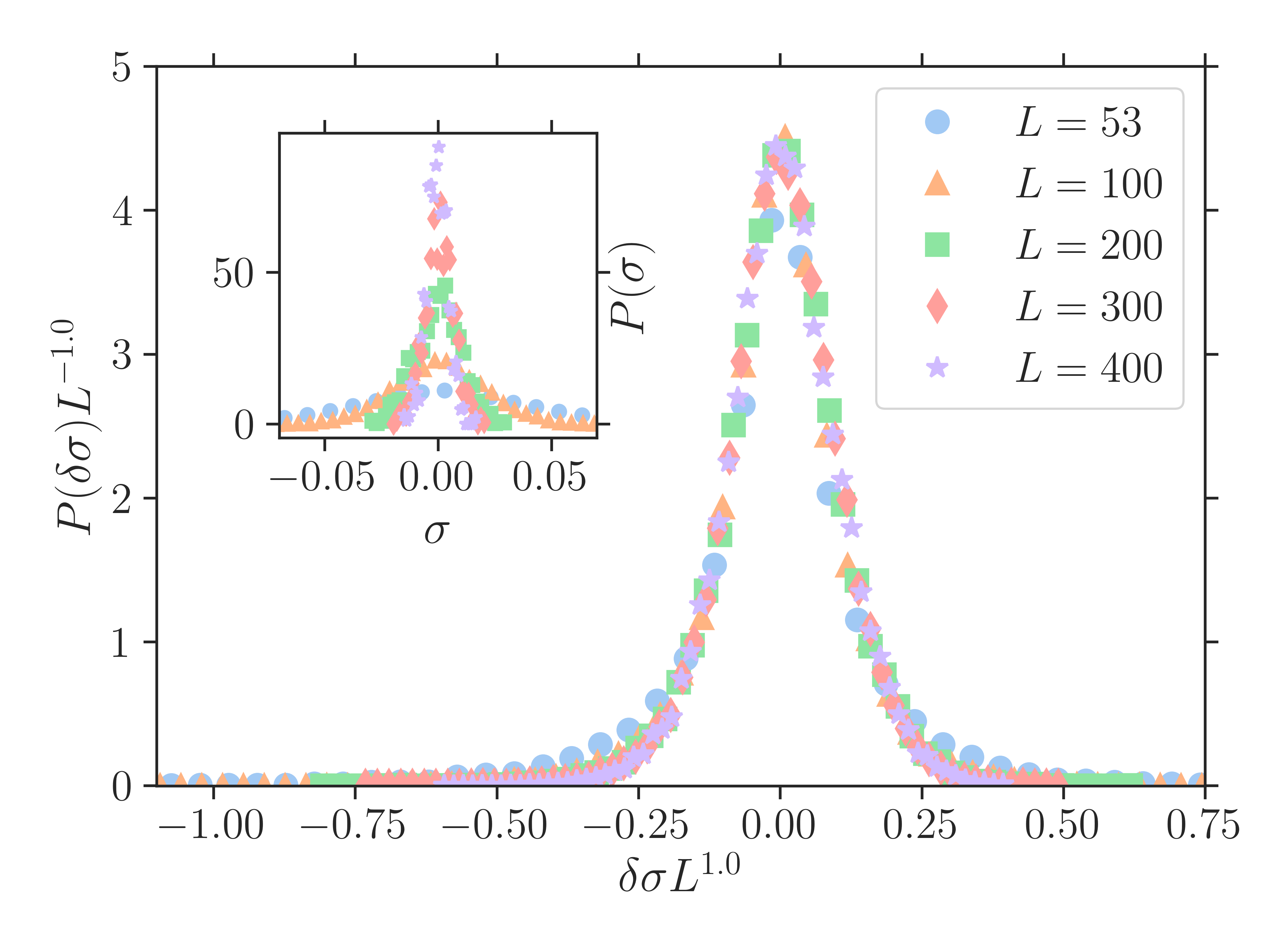}
\caption{Rescaled fluctuations of the stress in the stationary (top), transient (middle) and elastic (bottom) regime. The inset shows the system size dependence of the value of stress measured before and after avalanches.}
\label{fig:5}       
\end{figure}

\subsection{Elastic regime}
In the elastic regime  (strain interval 0-1\%), we see in Fig.~\ref{fig:7} (top) that the region in the scaled avalanche distributions that could be described by a power law has shrunk below one decade. This limited region would be consistent with an avalanche exponent \textcolor{black}{$\tau=1.00 \pm 0.06$}. Additionally, the fractal dimension has further decreased in comparison to the transient and is now \textcolor{black}{$d_f=0.51 \pm 0.25$.} 

The energy drop distributions by contrast now exhibit continuous curvature in the log-log representation, and therefore can no longer described by a power law. The inset shows that the avalanche size varies sublinearly with the energy drop over the entire range of values. Our data contrasts that of Shang and Barrat \cite{ShangBarrat}, who obtained much better defined power laws with exponents near 1.0 in the elastic regime for energy drops using the same 2D glass model and AQS protocol. They interpreted the value of $\tau=1.0$ as evidence for the marginally stable nature of the freshly quenched glass. An important difference between their study and ours is, however, that our glasses were prepared through a continuous quench to zero temperature, while Shang and Barrat used an \textcolor{black}{instantaneous} quench from the equilibrium state. It thus appears that the degree of marginality in Lennard Jones glasses is very sensitive to the preparation protocol as suggested by Scalliet {\it{et al.}} \cite{scalliet2019}. 

\subsection{Correlation length}
The above results show the importance of the spatial extent of avalanches for the shape of their distribution. To get more insight into the spatial correlations of stress in our system, we adopt the approach of Salerno and Robbins \cite{SalernoRobbins2013} and look at the system size scaling of the stress fluctuations. To this aim, we collect the value of the stress $\sigma$ right before and right after an avalanche and compute its distribution. In the inset of Figure \ref{fig:5} (top), we see that, as expected for the stationary regime, the stress fluctuates around its average value and the larger $L$ the more peaked is the distribution. The system size dependence of the distribution is given by
\begin{align}
P(\sigma) = L^{\phi} f(\delta \sigma L^{\phi})	
\label{sigmascaling}
\end{align}
with $\delta \sigma = \sigma - \langle \sigma \rangle_L$. The subscript $L$ indicates that the average is performed at a given system size. 
In the main panel of Figure \ref{fig:5} (top) we show the rescaled distribution of stress fluctuations. A good collapse is obtained for $\phi=0.85$, which is below the expected value of $\phi =1.0$ found for overdamped dynamics in two dimensions \cite{SalernoRobbins2013}, but again much closer to the value of $\phi = 0.9$ reported for critical damping and also quite compatible with the value $\phi=0.8$ reported by Lerner and Procaccia in an earlier study based on the same AQS scheme used here \cite{LernerProcaccia2009}. As $\phi \ne d/2$ we conclude that spatial correlations of the stress are present in our system. As discussed in ref \cite{SalernoRobbins2013}, the fluctuations of the stress with respect to its critical values set a correlation length $\xi \sim | \sigma_c - \sigma|^{-\nu}$ where $\nu=1/\phi=1.17$ in agreement with results from elastoplastic models \cite{LinWyartPNAS}. 

The distributions of the stress in the transient and elastic regimes are shown in Figure \ref{fig:5} (middle and bottom). Here, the stress fluctuations are Gaussian and their distributions collapse for $\phi=d/2=1$. In contrast to the steady state flow regime, this behavior suggests incoherent addition of stress coming from independent regions. Close proximity to criticality is thus needed to build up correlated behavior. 
\begin{figure}
 \includegraphics[scale=0.5]{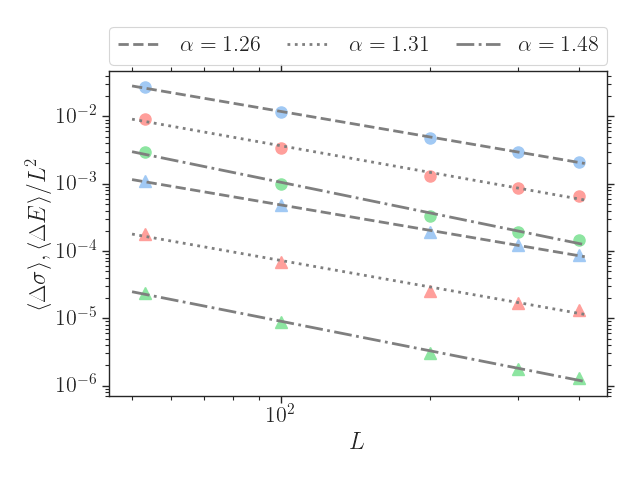}
 \caption{Scaling relations of the average stress drops ($\circ$) (blue) and energy drops ($\triangle$) for the stationary regime (blue), transient regime (red) and the elastic regime (green). }
\label{fig:6}       
\end{figure}

\subsection{Finite size scaling of the mean avalanche size}
The system size scaling of the mean stress drops $\langle \Delta \sigma \rangle \sim L^{-\alpha}$ is a further important characteristic of the athermal yielding transition. According to Fig.~\ref{fig:6}, the exponent $\alpha$ is 1.26, 1.31 and 1.48 for steady state, transient and elastic regime, resp. These values are in close agreement with the values 1.32,1.35, and 1.58 predicted from the scaling relation (\ref{alphatau}) for these regimes. Our steady state value $\alpha=1.26$ is in fact identical to that obtained by Lerner and Procaccia using a similar 2D model and protocol \cite{LernerProcaccia2009}. This value is in excellent agreement with the value $\alpha=4/3$ that was proposed as exact based on the singular behavior of the eigenvalues of the dynamical matrix near mechanical instability \cite{Karmakar2010Rapid}. The energy drops per unit volume exhibit the same finite size scaling in all regimes. This must be so because on average energy and stress drops coincide \cite{LernerProcaccia2009}. The first moments of their respective distributions are thus the same although the full distributions differ in general.

\begin{table*}[t]
\caption{Summary of the different exponents for the transient and stationary regimes obtained in this work. When two values are indicated, they correspond to the total (precursor) values of the considered exponent. \textcolor{black}{Values extracted from fits are indicated in the first 4 columns and values of $\tau$ and $\theta$ estimated from scaling relations are shown in the last 3 columns.}}
\label{tab:1}      
\begin{tabular}{|c|c|c|c|c|c|c|c|}
\hline\noalign{\smallskip}
Regime & $\tau $   & $d_f$ & $\phi$ & $\alpha$ & $\alpha$ (eq. \ref{alphatau})  &  $\theta$ (eq. \ref{tauLW}) &  $\theta$ (eq. \ref{tauKRR}) \\
\noalign{\smallskip}\hline\noalign{\smallskip}
Elastic  & $1.00$   & $0.51$  & $1.0$ & $1.48$ & $1.49$ &  $0.34$  & $0.34$  \\
Transient  & $0.86$ \, ($0.87$)  & $0.77$  \, ($0.57$) & $1.0$ & $1.31$ & 1.23 & $0.78\, (0.48)$ &  $0.71\, (0.46) $   \\
Stationary & $0.98$ \, ($1.12$)  & $0.88$ \, ($0.64$) & $0.85$ & $1.26$ & $1.14$ & $0.81\, (0.39)$ &  $0.80\, (0.41)$ \\ 
\noalign{\smallskip}\hline
\end{tabular}
\end{table*}

\section{Discussion}
The scaling exponents obtained in this work are summarized in Table \ref{tab:1}. 
The present study reveals that care must be taken with the characterization of the critical behavior of the yielding transition through atomistic simulations. When stress drops are used to define avalanches, our data suggests that the power law behavior becomes better defined as plastic activity increases. The fractal dimension $d_f$ increases strongly while the avalanche exponent $\tau$ remains nearly constant as the critical flow stress is approached. Ozawa {\it{et al.}} also reported no change between transient and stationary regime in the value of $\tau$ for a 3D polydisperse system \cite{OzawaPNAS}. Most of the scale free behavior of $P(S)$ is due to precursor events, and the measured values of $\tau$ are lower when mainshocks are included in the distribution. The increase of $d_f$ reflects a change in shape of the plastic events from point-like $(d_f=0)$ to more line-like $(d_f=1)$. Stress fluctuations only become correlated when the stress reaches the critical flow stress. The marginality exponent $\theta$ also slightly increases with loading. In previous work, we investigated the deformation of the quenched state up to the first plastic events. From the distribution of weakest sites $P(x_{min})$  \cite{Karmakar2010Rapid}, we found that $\theta \approx 0.37$ for the same model systems studied here \cite{RuscherRottler2020}. Here, we consider all events in interval 0-1$\%$, and the exponent $\theta=0.34$ is slightly lower than the one for first events in agreement with what has been found in other work \cite{Hentschel2015,ShangBarrat,RuscherRottler2020}.

While measuring the exponents $\tau$ and $d_f$ requires the full avalanche size distribution, determining the exponent $\alpha$ is easier since it requires only knowledge on their first moment. Our data suggests that the variation of $\alpha$ with strain is primarily driven by a change of avalanche shape. Moreover, the mean energy and stress drops are equivalent (Fig.~\ref{fig:6}). However, these quantities are not equivalent at the level of individual avalanches. In the pre-yield regimes, stress and energy drops are not proportional in the present Lennard-Jones glass former. A regime of proportionality only appears in steady state for large avalanches. This should suggest that at least in this regime their distributions should eventually become proportional as well. The system sizes required to access this regime cannot be studied with typical computational resources.  

With several similar studies of the subject now available in the literature, it has become clear that details of the model glass also affect the results. This can be most dramatically seen in the difference of the steady-state precursor values of  exponent $\tau$ between our work (1.12) and that of Oyama et al. \cite{Oyama2020} (1.5). While both studies used a 2D Lennard-Jones glass, ours included the attractive tail of the potential while that of Oyama et al. \cite{Oyama2020} had almost repulsive interactions. It appears that details of the local packing lead to small quantitative differences in the avalanche statistics. As expected, elastic and transient regime are sensitive to the glass preparation history, and the observation of a marginally stable phase in Lennard-Jones glass formers depends strongly on the choice of protocols and potential parameters.

\begin{acknowledgements}
Computing resources were provided by Compute Canada and by the HPC resources of IDRIS under the allocation 2020-A0090912010 made by GENCI. C.R acknowledges financial support by the ANR LatexDry project grant ANR-18-CE06-0001 of the French Agence Nationale de la Recherche. 

\end{acknowledgements}

%
\section*{Declarations}
\subsection*{Funding:} Computing resources were provided by Compute Canada and by the HPC resources of IDRIS under the allocation 2020-A0090912010 made by GENCI. C.R acknowledges financial support by the ANR LatexDry project grant ANR-18-CE06-0001 of the French Agence Nationale de la Recherche. 
\subsection*{Conflict of interest:} The authors declare that they have no conflict of interest.
\subsection*{Availability of data and material:} The raw simulation data can be obtained from the authors upon request.
\subsection*{Code availability:} The LAMMPS code (\texttt{https://lammps.sandia.gov/}) was used for all simulations in this work.

\bibliographystyle{spphys}       
\bibliography{biblio.bib}   


\end{document}